\documentstyle[prd,aps]{revtex}

\input epsf

 \mathchardef\ScriptA="7241
 \mathchardef\ScriptB="7242 
 \mathchardef\ScriptC="7243
 \mathchardef\ScriptD="7244
 \mathchardef\ScriptE="7245
 \mathchardef\ScriptF="7246
 \mathchardef\ScriptG="7247
 \mathchardef\ScriptH="7248
 \mathchardef\ScriptI="7249
 \mathchardef\ScriptJ="724A
 \mathchardef\ScriptK="724B
 \mathchardef\ScriptL="724C
 \mathchardef\ScriptM="724D
 \mathchardef\ScriptN="724E
 \mathchardef\ScriptO="724F
 \mathchardef\ScriptP="7250
 \mathchardef\ScriptQ="7251
 \mathchardef\ScriptR="7252
 \mathchardef\ScriptS="7253
 \mathchardef\ScriptT="7254
 \mathchardef\ScriptU="7255
 \mathchardef\ScriptV="7256
 \mathchardef\ScriptW="7257
 \mathchardef\ScriptX="7258
 \mathchardef\ScriptY="7259
 \mathchardef\ScriptZ="725A

\def\mapgeq{\mathop{>}\limits_{\sim}}
\def\mapleq{\mathop{<}\limits_{\sim}}

 \mathchardef\#="0023
 \mathchardef\$="0024
 \mathchardef\%="0025
 \mathchardef\ddash="705C
 
 \mathchardef\lwavy="336E
 \mathchardef\rwavy="336F
 \mathchardef\biglwavy="331A
 \mathchardef\bigrwavy="331B
 \mathchardef\bigglwavy="3328
 \mathchardef\biggrwavy="3329
 \mathchardef\littlesum="0350

\tighten
\draft
\begin{document} 
\bibliographystyle{prsty}

\title{
Neutrino Oscillations Induced by Two-loop Radiative Mechanism
}

\author{
Teruyuki Kitabayashi$^a$
\footnote{E-mail:teruyuki@post.kek.jp}
and Masaki Yasu${\grave {\rm e}}^b$
\footnote{E-mail:yasue@keyaki.cc.u-tokai.ac.jp}
}

\address{\vspace{5mm}$^a${\sl Accelerator Engineering Center} \\
{\sl Mitsubishi Electric System \& Service Engineering Co.Ltd.} \\
{\sl 2-8-8 Umezono, Tsukuba, Ibaraki 305-0045, Japan}}
\address{\vspace{2mm}$^b${\sl Department of Natural Science\\School of Marine
Science and Technology, Tokai University}\\
{\sl 3-20-1 Orido, Shimizu, Shizuoka 424-8610, Japan}\\and\\
{\sl Department of Physics, Tokai University} \\
{\sl 1117 KitaKaname, Hiratsuka, Kanagawa 259-1292, Japan}}
\date{TOKAI-HEP/TH-0001, June, 2000}
\maketitle

\begin{abstract}
Two-loop radiative mechanism, when combined with an $U(1)_{L^\prime}$ symmetry generated by 
$L_e$ $-$ $L_\mu$ $-$ $L_\tau$ (=$L^\prime$), is shown to provide an estimate of 
$\Delta m^2_{\odot}$/$\Delta m^2_{atm}$ $\sim$ $\epsilon m_e/m_\tau$, 
where $\epsilon$ measures the 
$U(1)_{L^\prime}$-breaking.  Since $\Delta m^2_{atm}$ $\sim$ 3.5$\times 10^{-3}$ eV$^2$, we find that 
$\Delta m^2_{\odot}$ $\sim$ $\epsilon$10$^{-6}$ eV$^2$, which will fall into 
the allowed region of the LOW solution to the solar neutrino problem for $\epsilon$ $\sim$ 0.1.
\end{abstract}
\vspace{2mm}
Recent evidence for atmospheric neutrino oscillations \cite{SuperKamiokande} has promoted new 
theoretical activities of understanding properties of neutrinos, especially concerning the 
long-standing theoretical issue of their masses and mixings \cite{EarlyMassive}.  It is known that 
there have been so far two main ideas to account for the smallness of 
neutrino masses, which are, respectively, called as seesaw mechanism \cite{SeeSaw} and 
as radiative mechanism \cite{Zee,Babu}. Various possibilities of realizing the radiative 
mechanism in neutrino physics have been discussed \cite{Radiative,ZeeType,Usefulness}.  
Especially, in recent analyses on neutrino mass matrix of the Zee-type \cite{Zee}, 
the usefulness of the conserved quantum number of $L_e$ $-$ $L_\mu$ $-$ $L_\tau$ 
(=$L^\prime$) \cite{LeLmuLtau} has been recognized \cite{Usefulness}. This $U(1)_{L^\prime}$ 
symmetry works to yield maximal mixing in both atmospheric 
and solar neutrino oscillations \cite{BiMaximal}.  The maximal mixing in atmospheric neutrino oscillations has 
been supported by 
the data indicating that $\sin^2 2\theta_{23}$ $\sim$ 1 with $\Delta m^2_{23}$ $\sim$ 
$3.5 \times 10^{-3}$ eV$^2$ \cite{SuperKamiokande}, 
where $\theta_{ij}$ stands for the mixing angle and $\Delta m^2_{ij}$ stands for the squared mass 
difference for $\nu^i$ $\leftrightarrow$ $\nu^j$.  Solar neutrino oscillations have also been 
considered to exhibit 
the maximal mixing if the oscillations are described by 
$\Delta m^2_{12}$ $\sim$ $3 \times 10^{-5}$ eV$^2$ as a large mixing angle solution (LMA), 
$\sim$ $10^{-7}$ eV$^2$ as a less probable solution with low probability and low mass (LOW) 
and $\sim$ $10^{-10}$ eV$^2$ as a vacuum oscillation solution (VO) \cite{MSW}. 

In the present article, we further apply the ans{\"a}tz of 
the $L^\prime$-conservation to models of neutrino masses 
based on two-loop radiative mechanism \cite{Babu}. It is anticipated 
to provide more natural explanation of the tiny neutrino mass without enhanced suppression in 
couplings, which is experimentally 
of order 0.01 eV \cite{TwoLoopGood}.  Furthermore, some flavor-changing interactions receive 
extra suppression owing to the presence of the approximate $U(1)_{L^\prime}$ symmetry.
It should be also noticed that, in the radiative mechanism of the Zee type, which is based on 
one-loop diagrams, fine-tuning of lepton-number violating couplings is necessary to 
yield bimaximal mixing even if one invokes the $L^\prime$-conservation. The fine-tuning 
can be characterized by $\ddash$inverse hierarchy in the couplings", namely, 
$f_{[13]}m^2_\tau$ $\sim$ $f_{[12]}m^2_\mu$ \cite{Usefulness}, where $f$'s are to be defined in Eq.(1). 
In the present model, it will be shown that nearly bimaximal structure is dynamically guaranteed 
by the heaviness of the $\tau$ lepton and by the lightness of the electron.  Therefore, no fine-tuning is 
necessary.

The two-loop radiative mechanism can be 
embedded in the standard model by employing two $SU(2)_L$-singlet charged Higgs scalar, 
$h^+$ and $k^{++}$, in addition to the standard Higgs, $\phi$.  
The extra Higgses, $h^+$ and $k^{++}$, 
respectively, couple to charged lepton-neutrino pairs and charged lepton-charged lepton pairs. 
Their interactions are described by 
\begin{eqnarray}
-{\cal L}_h  & = &  
\sum_{i,j=1,2,3}
\frac{1}{2}f_{[ij]}
{\overline {\left( \psi_L^i \right)^c}}\psi^j_Lh^+
+f_{\{ij\}}{\overline {\left( \ell^i_R \right)^c}}\ell^j_Rk^{++} 
+ {\rm (h.c.)},
\label{Eq:Yukawa}
\end{eqnarray}
where $\psi_L^i$ and $\ell^i_R$ ($i$=1,2,3) stand for three families of leptons and 
the Yukawa couplings, $f$'s, satisfy $f_{[ij]}$ = $-f_{[ji]}$ and  $f_{\{ij\}}$ = $f_{\{ji\}}$.  
Now, let us introduce the $U(1)_{L^\prime}$ symmetry into ${\cal L}_h$.  By envisioning 
the import of its breaking effect, we employ an additional $k^{++}$ to be denoted by $k^{\prime ++}$.  
The quantum number, $L^\prime$, is assigned to be 1 for ($\psi^1_L$, $\ell^1_R$), 0 for 
($\phi$, $h^+$, $k^{++}$), $-1$ for ($\psi^{2,3}_L$, $\ell^{2,3}_R$) and 
$-2$ for $k^{\prime ++}$. The ordinary lepton number, to be denoted by $L$, 
can also be assigned to be 1 for leptons, 0 for $\phi$ 
and $-2$ for ($h^+$, $k^{++}$, $k^{\prime ++}$).  

Yukawa interactions take the form of 
\begin{eqnarray}
-{\cal L}_Y  & = &  
\sum_{i=1,2,3}
f^i_\phi {\overline {\psi^i_L}}\phi\ell^i_R
+
\sum_{i=2,3}
\left(
\frac{1}{2}f_{[1i]}{\overline {\left( \psi_L^1 \right)^c}}\psi^i_Lh^+
+f_{\{1i\}}{\overline {\left( \ell^1_R \right)^c}}\ell^i_Rk^{++} 
\right)
\nonumber \\
& &+ 
 f_{\{11\}}{\overline {\left( \ell^1_R \right)^c}}\ell^1_Rk^{\prime ++} 
+ {\rm (h.c.)},
\label{Eq:OurYukawa}
\end{eqnarray}
and Higgs interactions are described by
self-Hermitian terms composed of $\varphi\varphi^\dagger$ ($\varphi$ = $\phi$, $h^+$, 
$k^{++}$, $k^{\prime ++}$) and by the non-self-Hermitian terms in
\begin{eqnarray}\label{Eq:Conserved}
V_0 & = &  
\mu_0h^+h^+k^{++^\dagger}
+ {\rm (h.c.)},
\end{eqnarray}
where $\mu_0$ represents a mass scale.  This coupling softly breaks the $L$-conservation 
but preserves the $L^\prime$-conservation.  
To account for solar neutrino oscillations, the breaking of the $L^\prime$-conservation 
should be included and is assumed to be furnished by 
\begin{eqnarray}\label{Eq:Broken}
V_b & = &\mu_bh^+h^+k^{\prime ++\dagger} + {\rm (h.c.)},
\end{eqnarray}
where $\mu_b$ represents a breaking scale of the $L^\prime$-conservation.  One can instead 
introduce a neutral Higgs scalar, which spontaneously breaks $U(1)_L$ (or $U(1)_{L^\prime}$) 
by acquiring vacuum expectation value related to $\mu_0$ (or $\mu_b$) \cite{Majoron}.  
However, there necessarily appears a Nambu-Goldstone boson called Majoron, whose coupling to 
matter should be kept sufficiently small. To avoid having such a dangerous massless Majoron 
is to include soft $U(1)$-breaking interactions such as Eqs.(\ref{Eq:Conserved}) and 
(\ref{Eq:Broken}), which generate its mass of order of the breaking mass scale.

Neutrino masses are generated by interactions corresponding to Fig.1 
for the $U(1)_{L^\prime}$-conserving processes and 
Fig.2 for the $U(1)_{L^\prime}$-breaking processes. 
The resulting neutrino mass matrix is found to be
\begin{eqnarray}\label{Eq:NUmass}
M_\nu = \left( \begin{array}{ccc}
  0&  m_{12}&   m_{13}\\
  m_{12}&  m^\prime_{22}&  m^\prime_{23}\\
  m_{13}&  m^\prime_{23}&  m^\prime_{33}\\
\end{array} \right),
\end{eqnarray}
where $m_{1i}$ and $m^\prime_{ij}$ ($i$,$j$ = 2,3) are calculated to be
\begin{eqnarray}
m_{1i}  &=&  2\sum_{j=2,3}f_{[1j]}f_{\{j1\}}f_{[1i]}
\frac{m_{\ell_j}m_e\mu_0}{m^2_k}F(m^2_{\ell_j}, m^2_h, m^2_k)F(m^2_e, m^2_h, m^2_k),  
\label{Eq:MatrixEntryIJ1} \\
m^\prime_{ij}  &=& 
-f_{[1i]}f_{\{11\}}f_{[1j]}
\frac{m_em_e\mu_b}{m^2_{k^\prime}}F(m^2_e, m^2_h, m^2_{k^\prime})F(m^2_e, m^2_h, m^2_{k^\prime})
\label{Eq:MatrixEntryIJ2}
\end{eqnarray}
under the approximation that $m^2_{k,k^\prime}$ $\gg$ $m^2_{\ell^i, e, h}$. 
Mass parameters, $m_{k, k^\prime, h}$, respectively, stand for the masses of 
Higgs scalars, $k^{++}$, $k^{\prime ++}$ and $h$, and the function of $F$ is 
defined by
\begin{eqnarray}
&&F(x, y, z) = \frac{1}{16\pi^2}
\frac{x\ln\left( x/z\right)-y\ln\left( y/z\right)}{x-y}.
\label{Eq:F_x_y}
\end{eqnarray}
The outline of its derivation can be seen from the Appendix.

The entries of $m_{12}$ and $m_{13}$ receive contributions from both $\mu$- and $\tau$-exchange 
as in Eq.(\ref{Eq:MatrixEntryIJ1}).  Since $m_\tau$ $\gg$ $m_\mu$, the $\tau$-exchange gives dominant 
contributions to $m_{12}$ and $m_{13}$, which result in the same mass-dependence.  One can, 
then, observe that $m_{12}$ $\sim$ $m_{13}$ is a natural consequence without fine-tuning of the couplings. 
In fact, nearly bimaximal mixing is reproduced by $f_{[12]}$ $\sim$ $f_{[13]}$. 
Other entries, $m^\prime$'s, are further suppressed by the factor of $m_e/m_\tau$.  
Thus, the form of our neutrino mass matrix is consistent with the one described by nearly bimaximal mixing.  

We find that $\Delta m^2_{atm}$ for atmospheric neutrino oscillations 
and $\Delta m^2_{\odot}$ for solar neutrino oscillations are calculated to be:
\begin{equation}\label{Eq:NUmass2}
\Delta m^2_{atm} = m^2_{12} + m^2_{13} (\equiv m^2_\nu), \ \ \
\Delta m^2_{\odot} = 4 m_\nu\delta m
\end{equation}
with
\begin{eqnarray}\label{Eq:TinyNU}
&\delta m = \frac{1}{2}\vert m^\prime_{22}\cos^2\theta_\nu  + 2m^\prime_{23}\cos\theta_\nu \sin\theta_\nu  
+ m^\prime_{33}\sin^2\theta_\nu \vert, 
\end{eqnarray}
where the mixing angle $\theta_\nu$ is defined by $\cos\theta_\nu$ = $m_{12}/m_\nu$ 
($\sin\theta_\nu$ = $m_{13}/m_\nu$) 
and the anticipated relation of $\vert m^\prime_{ij}\vert$ $\ll$ $\vert m_{1k}\vert$ 
for $i,j,k$ = 2,3 have been used.  As far as mass scales are concerned,
\footnote{
The possible 
contributions to the $\nu_e-\nu_e$ entry of $M_\nu$ in Eq.(\ref{Eq:NUmass}), arising from the 
three loop diagrams found in Ref.\cite{Same}, 
come from four-loop diagrams involving the loop for the $k^{++}$-$k^{\prime ++}$ mixing characterized by 
the factor of $\xi$ ($\sim$ $(16\pi^2)^{-1}(\mu_0/m_k)(\mu_b/m_{k^\prime})$). 
This diagram at most yields $\delta$ 
$\sim$ $(16\pi^2 )^{-1}\xi m_\tau^2/m^2_k$, which should be compared with $m_e^2/m^2_{k^\prime}$. 
Our estimate of $\Delta m^2_{\odot}$ $\propto$ $m^2_e$ is not drastically altered by 
including this effect 
since the present parameter set gives $\delta$ 
$\sim$ 2$m_e^2/m^2_k$, which turns out to be ${\cal O}$($m_e^2/m^2_{k^\prime}$).
}
we reach 
\begin{eqnarray}\label{Eq:OurRelation}
 \Delta m^2_{\odot} \sim \frac{\mu_b}{\mu_0}\frac{m_e}{m_\tau}\frac{m^2_k}{m^2_{k^\prime}}
\Delta m^2_{atm}.
\end{eqnarray}
It turns out to be $\Delta m^2_{\odot}$/$\Delta m^2_{atm}$ $\sim$ $\epsilon m_e/m_\tau$ for 
$m_k$ $\sim$ $m_{k^\prime}$, where $\epsilon$ $\sim$ $\mu_b/\mu_0$, which is the announced result. 
The experimental value of 
$\Delta m^2_{atm}$ $\sim$ $3.5\times10^{-3}$ eV$^2$  \cite{SuperKamiokande} 
implies $\Delta m^2_{\odot}$ $\sim$ $10^{-7}$ eV$^2$ for $\epsilon$ $\sim$ 0.1,
which lies in the region corresponding to the LOW solution to the solar 
neutrino problem \cite{MSW}. 

To see order of magnitude estimates of our parameters, we have to first recognize possible constraints 
on masses and couplings 
since the interactions mediated by $h^+$, $k^{++}$ and $k^{\prime ++}$ disturb the well established 
low-energy phenomenology.  The most stringent constraints on $f$'s, which are relevant for 
our discussions, are listed as
\begin{eqnarray}\label{Eq:Constraint1}
&& \xi\vert\frac{f_{\{11\}}f_{\{12\}}}{{\bar m}^2_k}\vert 
< 2.9 \times 10^{-11} ~(6 \times 10^{-9})~{\rm GeV}^{-2} ,
\end{eqnarray}
from $\mu^-$ $\rightarrow$ $e^-e^-e^+$ with 
BR($\mu^-$ $\rightarrow$ $e^-e^-e^+$) $<$ 10$^{-12}$ \cite{Data}
($\mu^-$ $\rightarrow$ $e^-$ + $\gamma$ with BR($\mu^-$ $\rightarrow$ $e^-$ + $\gamma$) $<$ 
4.9$\times 10^{-11}$  \cite{Data}) \cite{Mu_E}, 
where $\xi$ $\sim$ $(16\pi^2)^{-1}(\mu_0/m_k)(\mu_b/m_{k^\prime})$ 
arising from the loop for the $k^{++}$-$k^{\prime ++}$ mixing, which 
represents the extra 
suppression factor due to $U(1)_{L^\prime}$ and 
${\bar m}^2_k$ stands for the averaged mass of $k^{++}$ and $k^{\prime ++}$, 
and 
\begin{eqnarray}\label{Eq:Constraint2}
&& \vert\frac{f_{\{11\}}}{m_{k^\prime}}\vert^2 < 1.2\times 10^{-5} ~{\rm GeV}^{-2},
\end{eqnarray}
from $e^-e^-$ $\rightarrow$ $e^-e^-$ \cite{E_E}.  The contributions to this process via the $k^{++}$-exchanges 
turn out to be higher loop-effects since $k^{++}$ does not directly couple to $e^-e^-$ and are 
expected to be well suppressed.  The $\mu$ decay of 
$\mu^-$ $\rightarrow$ $\nu_\mu e^-{\bar \nu_e}$ is used to 
determine the value of the Fermi constant, which includes the extra $h^+$-contributions, 
thus, providing slight deviation of the electroweak 
gauge coupling of $g$ from the standard value; therefore, the constraint should 
be deduced from that on $g$ \cite{MuDecay}, which can be translated into
\begin{eqnarray}\label{Eq:Constraint3}
&& \vert\frac{f_{[12]}}{m_h}\vert^2 < 1.7 \times 10^{-6} ~{\rm GeV}^{-2},
\end{eqnarray}
for $\nu_\mu({\bar \nu_\mu})e^-\rightarrow \nu_\mu({\bar \nu_\mu})e^-$.

For the present analysis, the couplings of $f$'s are kept as small as 
${\cal O}$($e$).  We adopt the following parameter values that satisfy 
these constraints, where $f_{[12]}$ $\sim$ $f_{[13]}$ is assumed 
to yield nearly bimaximal mixing: 
$f_{[12]}$ $\sim$ $f_{[13]}$ $\sim$ 2$e$ 
yielding $m_h$ $\mapgeq$ 350 GeV by Eq.(\ref{Eq:Constraint3}), 
from which $m_h$ $\sim$ 350 GeV is taken, 
$f_{\{11\}}$ $\sim$ $f_{\{13\}}$ $\sim$ $e$, $m_k$ $\sim$ 2 TeV 
with $m_k-m_{k^\prime}$ $\sim$ $m_k$/10 
and $\mu_0$ $\sim$ 1.5 TeV with $\mu_b$ $\sim$ $\mu_0$/10 giving $\epsilon$ $\sim$ 0.1.
The constraint of Eq.(\ref{Eq:Constraint1}) is satisfied by $f_{\{12\}}$ $\mapleq$ 1. 
These parameters, in fact, reproduce $\Delta m^2_{atm}$ $\sim$ 
$2.4\times 10^{-3}$ eV$^2$ and $\Delta m^2_{\odot}$ $\sim$ $10^{-7}$ eV$^2$, 
which is relevant for the LOW solution to the solar neutrino problem.
\footnote{
Of course, $\epsilon$ $\sim$ 10$^{-4}$ gives $\Delta m^2_{\odot}$ $\sim$ $10^{-10}$ eV$^2$, 
corresponding to the VO solution \cite{Same}.  However, it is not suitable for 
our discussions to obtain a tiny mass-splitting without such enhanced suppression in couplings.
}
The mass scale of 
the heaviest neutrino mass is characterized by $(16\pi^2)^{-2}(m_em_\tau\mu_0/m^2_k)$ $\sim$ 0.01 eV.

To conclude, we have demonstrated that two-loop radiative mechanism well works to account for neutrino 
oscillation phenomena 
when it is combined with the $U(1)_{L^\prime}$ symmetry.  Thanks to the well known loop-factor of 
$(16\pi^2)^{-2}$, neutrino masses are well suppressed to yield ${\cal O}$(0.01) eV.  The couplings of $f$'s 
can be chosen to be ${\cal O}(e)$ as $f_{[12]}$ $\sim$ $f_{[13]}$ $\sim$ 2$e$ and 
$f_{\{11\}}$ $\sim$ $f_{\{13\}}$ $\sim$ $e$. 
Solar neutrino oscillations are controlled by the factor of $(m_e/m_{k^\prime})^2$, leading to the relation of 
$\Delta m^2_{\odot}$ $\sim$ $(m_e\mu_b/m_\tau\mu_0)\Delta m^2_{atm}$,
which provides the LOW solution for $\mu_b$ $\sim$ 0.1$\mu_0$.

The work of M.Y. is supported by the Grant-in-Aid for Scientific Research No 12047223 
from the Ministry of Education, Science, Sports and Culture, Japan.

Note added: While preparing this manuscript, we are aware of the article \cite{Same} that 
has treated the same subject and has reached slightly different conclusion.

\noindent
{\bf \center Appendix\\}
In this Appendix, we describe the outline of obtaining the integral of Eq.(\ref{Eq:F_x_y}) used in 
Eqs.(\ref{Eq:MatrixEntryIJ1}) and (\ref{Eq:MatrixEntryIJ2}).  From the diagram in Fig.1, we write 
the relevant integration to be:
\begin{eqnarray}\label{Eq:Start}
&&I = \int {\frac{{d^4 k}}{{\left( {2\pi } \right)^{4} }}} \frac{{d^4 q}}{{\left( {2\pi } \right)^{4} }}\frac{1}{{\left( {k^2  - m_\ell^2 } \right)\left( {k^2  - m_h^2 } \right)\left( {q^2  - m_e^2 } \right)\left( {q^2  - m_h^2 } \right)\left( {\left( {k - q} \right)^2  - m_k^2 } \right)}}.
\end{eqnarray}
By performing the intgration over $k$ supplemented by
\begin{eqnarray}\label{Eq:Feymann}
&& \frac{1}{abc} = \frac{{\Gamma \left( 3 \right)}}{{\Gamma \left( 1 \right)\Gamma \left( 1 \right)\Gamma \left( 1 \right)}}\int_0^1dx\int_0^1ydy \frac{1}{{\left[ {c + \left( {b - c} \right)y + \left( {a - b} \right)xy} \right]^3 }},
\end{eqnarray}
and by noticing the formula for the one-loop integral
\begin{eqnarray}\label{Eq:OneLoopIntegral}
&&\int \frac{d^4 q}{\left( {2\pi } \right)^{4} }\frac{1}{\left( q^2  - a\right)\left( q^2  - b \right)\left( q^2  - c  \right)} = - \frac{i}{16\pi ^2 }\left[ \frac{a \ln a }{\left( a-b\right)\left( a-c \right)} + \frac{b \ln b }{\left( b  - a  \right)\left( b  - c \right)} + \frac{c \ln c}{\left( c - a \right)\left( c  - b \right)} \right],
\end{eqnarray}
we reach 
\begin{eqnarray}\label{Eq:Result}
&&I = \int {dxydy\frac{i}{{16\pi ^2 \left[ y\left( {1 - y} \right) \right]}}} I\left( {x,y} \right),
\end{eqnarray}
where
\begin{eqnarray}\label{Eq:Details}
&& I\left( x,y \right) =  - \frac{i}{16\pi ^2 }\left[ \frac{m_e^2 \ln m_e^2 }{\left( {m_e^2  - m_h^2 } \right)\left( {m_e^2  - M^2 } \right)} + \frac{m_h^2 \ln m_h^2 }{\left( {m_h^2  - m_e^2 } \right)\left( {m_h^2  - M^2 } \right)}
+ \frac{M^2 \ln M^2 }{\left( {M^2  - m_h^2 } \right)\left( {M^2  - m_e^2 } \right)} \right]
\end{eqnarray}
with
\begin{eqnarray}\label{Eq:xyMass}
&&M^2  = \frac{m_k^2  - \left( m_k^2  - m_h^2 \right)y - \left( m_h^2  - m_\ell^2 \right)xy}{y\left( 1 - y \right)}.
\end{eqnarray}
Under the approximation of $m_k^2$ $\gg$ $m_{\ell, e, h}^2$, we find that
\begin{eqnarray}\label{Eq:Factorixation}
&&y\left( 1-y \right) \left( a - M^2 \right) \approx -a\left(y-\alpha \right)\left( y-\beta \right)
\end{eqnarray}
with 
\begin{eqnarray}\label{Eq:TwoRoots}
&&\alpha = \frac{m^2_k}{a}\left[ 1-\frac{m_h^2-\left( m_h^2-m_\ell^2 \right) x}{m_k^2}\right], \ \ \
\beta = 1 + \frac{m_h^2-\left( m_h^2-m_\ell^2 \right) x}{m_k^2},
\end{eqnarray}
which yield
\begin{eqnarray}\label{Eq:xyPerformed1}
&& J\left( a \right) = \int dxydy\frac{1}{ y\left( 1 - y \right)\left(a  - M^2 \right)}  
\approx  \frac{1}{m_k^2 }
\frac{m_\ell^2 \ln \left( m_\ell^2 / m_k^2\right)-m_h^2 \ln \left( m_h^2 / m_k^2\right)}
{m_\ell^2  - m_h^2 }
\end{eqnarray}
and
\begin{eqnarray}\label{Eq:xyPerformed2}
&&\int dxydy\frac{M^2\ln M^2}{ y\left( 1 - y \right)\left(a  - M^2 \right)\left(b  - M^2 \right)}  
\approx  - \frac{aJ\left( a \right) - bJ\left( b \right)}{a-b}\ln m_k^2,
\end{eqnarray}
where we have used $\ln M^2$ $\approx$ $\ln m_k^2$.  The parameters of $a$ and $b$ should satisfy the 
condition of $a,b$ $\ll$ $m_k^2$. 
The function of $J\left( a \right)$ turns out to 
be independent of $a$ in the present approximation. Collecting these results, we finally obtain
\begin{eqnarray}\label{Eq:FinalResult}
I = \frac{{F\left( {m_\ell^2 ,m_h^2 ,m_k^2 } \right)F\left( {m_e^2 ,m_h^2 ,m_k^2 } \right)}}{{m_k^2 }},
\end{eqnarray}
where
\begin{eqnarray}\label{Eq:FinalF_x_y}
F\left( x,y,z \right) = \frac{1}{16\pi ^2 }\frac{x\ln \left( x/z \right) - 
y\ln \left( y/ z\right)}{x-y},
\end{eqnarray}
which is the expression of Eq.(\ref{Eq:F_x_y}).


\noindent
{\bf \center Figure Captions\\}
\begin{description}
\item {\bf Fig.1} : $U(1)_{L^\prime}$-conserving two loop radiative diagrams for $\nu^1$-$\nu^i$ ($i$=2,3) via 
$\mu^-$ ($i$=2) and $\tau^-$ ($i$=3).
\item {\bf Fig.2} : $U(1)_{L^\prime}$-breaking two loop radiative diagrams for $\nu^i$-$\nu^j$  ($i,j$=2,3) via $e^-$.
\end{description}

\newpage
\epsfxsize=470pt
\epsfbox[90 0 530 750]{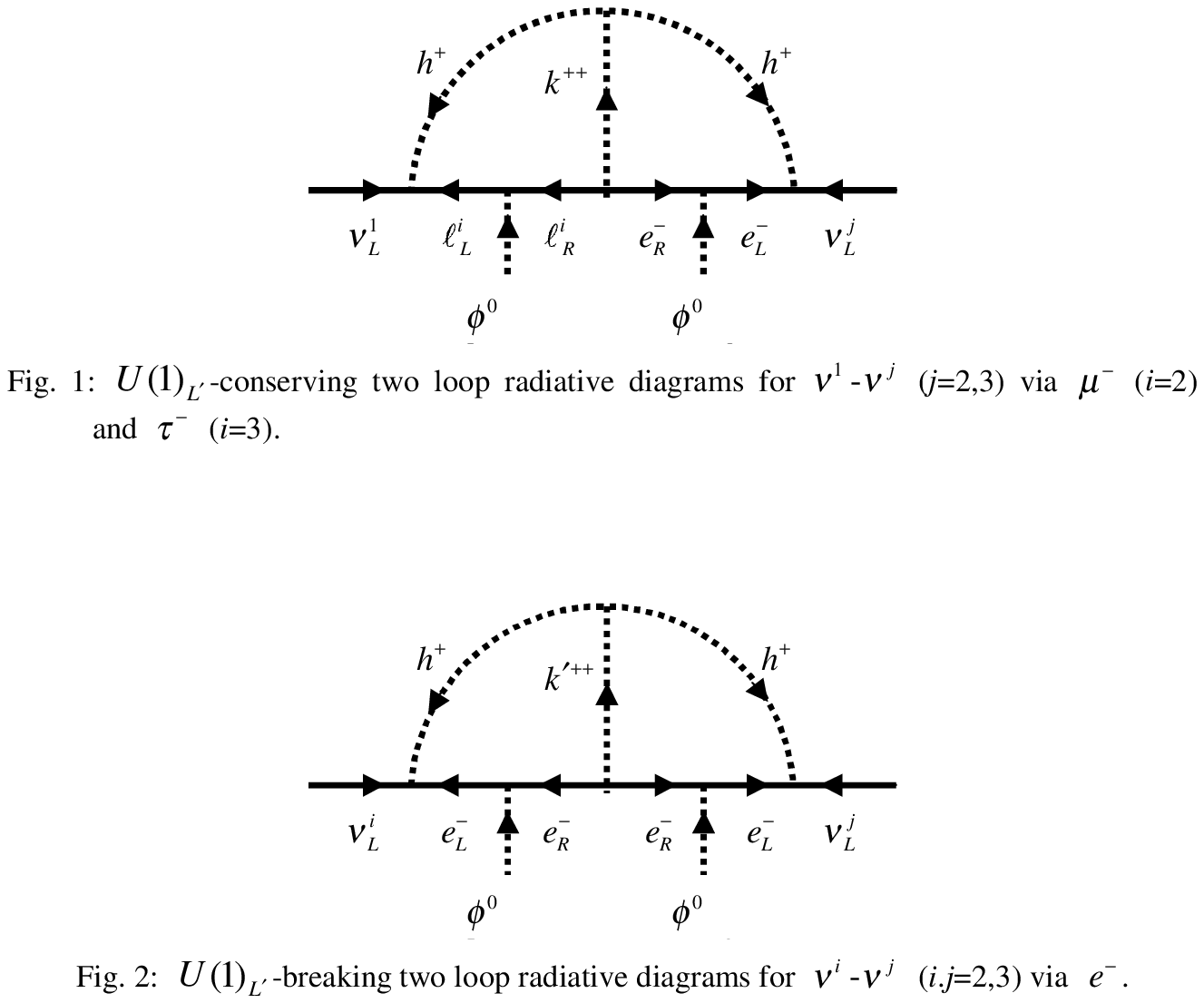}

\end{document}